\documentclass{ws-ijqi}
\usepackage[super,sort,compress]{cite}

\usepackage{}\newcounter{myctr}
\def\myitem{\refstepcounter{myctr}\bibfont\noindent\ifnum\themyctr>9\else\phantom{0}\fi\hangindent17pt\themyctr.\enskip}

\usepackage{graphicx,amsmath,braket,relsize,epstopdf,color,mathtools,bm,newtxtext,newtxmath,braket,rotating,upgreek,enumitem}
\usepackage[hyphenbreaks]{breakurl}
\usepackage[colorlinks=true,linkcolor=blue,citecolor=blue,urlcolor =blue]{hyperref}
\usepackage[normalem]{ulem}
\usepackage{comment}

\usepackage[table,xcdraw]{xcolor}

\newcommand{\Cov}{\mathop{\mathrm{Cov}} \nolimits}
\newcommand{\RE}{\mathop{\mathrm{Re}} \nolimits}
\newcommand{\IM}{\mathop{\mathrm{Im}} \nolimits}
\newcommand{\Tr}{\mathop{\mathrm{Tr}} \nolimits}
\newcommand{\iu}{\mathrm{i}}
\newcommand{\eu}{\mathrm{e}}
\newcommand{\op}[1]{\hat{#1}}
\newcommand{\D}{\mathrm{d}}
\newcommand{\bmtheta}{\bm{\uptheta}}

\begin{document}

\catchline{}{}{}{}{}

\title{Taming singularities of the quantum Fisher information}

\author{Aaron~Z.~Goldberg}
\address{National Research Council of Canada, 100 Sussex Drive\\
Ottawa, Ontario K1A 0R6, Canada\\
aaron.goldberg@nrc-cnrc.gc.ca}

\author{Jos\'e~L.~Romero$^{\ast}$ and \'Angel~S.~Sanz$^{\dag}$}
\address{Departamento de \'Optica, Facultad de F\'{\i}sica\\
Universidad Complutense, 28040 Madrid\\
\textbf{}$^{\ast}$joselr04@ucm.es \\
$^{\dag}$a.s.sanz@fis.ucm.es}

\author{Luis~L. S\'{a}nchez-Soto}
\address{Departamento de \'Optica, Facultad de F\'{\i}sica\\
Universidad Complutense, 28040 Madrid}
\address{Max-Planck-Institute f\"{u}r die Physik des Lichts\\
Staudtstrasse 2, 91058 Erlangen, Germany\\lsanchez@fis.ucm.es}

\maketitle

\begin{history}
\received{12 August 2021}
\revised{Day Month Year}
\end{history}

\begin{abstract}
Quantum Fisher information matrices (QFIMs) are fundamental to estimation theory: they encode the ultimate limit for the sensitivity with which a set of parameters can be estimated using a given probe. Since the limit invokes the inverse of a QFIM, an immediate question is what to do with singular QFIMs. Moreover, the QFIM may be discontinuous, forcing one away from the paradigm of regular statistical models. These questions of nonregular quantum statistical models are present in both single- and multiparameter estimation
. Geometrically, singular QFIMs  occur when the curvature of the metric vanishes in one or more directions in the space of probability distributions, while QFIMs have discontinuities when the density matrix has parameter-dependent rank.
We present a nuanced discussion of how to deal with each of these scenarios, stressing the physical implications of singular QFIMs and the ensuing ramifications for quantum metrology.
\end{abstract}



\markboth{Goldberg, Romero, Sanz, S\'anchez-Soto}
{Singular quantum Fisher information}

\section{Introduction}

High-precision measurements constitute one of the main driving forces for modern technology. However, physical laws dictate fundamental limits to the attainable precision, depending on the quantity being measured.  Still, most of the current limitations are not fundamental: the exploitation of quantum effects promises to overcome the limits that can be obtained by using only classical resources. This is the basis of the flourishing area of quantum metrology, which aims to determine and attain the ultimate bounds on measurement precision.   

Practical applications typically involve the determination of multiple parameters. The ultimate precision with which these parameters can be estimated in an ideal protocol is lower-bounded by the inverse of a quantity known as the quantum Fisher information matrix (QFIM)~\cite{Helstrom:1976ij}, which governs the sensitivity of a probe state to changes in the underlying parameters being estimated. This presents a serious challenge when the QFIM becomes singular~\cite{Tsang:2020ve}: can the lowest possible uncertainty for estimating a parameter really be infinite? Similarly, when the QFIM is discontinuous, can this prevent us from finding a measurement protocol that garners as much information as the QFIM? The answers depend on the situation at hand, as we will explore in detail.

Instances where QFIMs become singular abound: in estimating the parameters of a pure qubit, the coordinate system can diverge~\cite{Rafal:2020aa}; in estimating the parameters of a rotation, SU(2)-coherent states cannot distinguish between all of the free parameters~\cite{Goldberg:2021uw}; in absorption measurements, coherent states can fail in the limit of small losses~\cite{Monras:2007wx}; in superresolution of incoherent sources, image plane intensity measurements preclude measuring of the distance between two sources when it approaches zero~\cite{Tsang:2019ww}; and, in multiphase estimation without an external reference mode, one relative phase becomes immeasurable~\cite{Goldberg:2020vw}. This panoply of examples provides ample motivation for a comprehensive study of the underpinnings and implications of singular QFIMs.

From a geometrical point of view, the QFIM is intrinsically related to the Bures metric~\cite{Bures:1969aa}. This means that it governs the curvature in the space of probability distributions, allowing one to define distances between distributions and to compare how they change with changes in the underlying parameters. Metric singularities can readily be visualized as points of vanishing curvature, which aids in taming the deleterious effects of the singularities. Discontinuous QFIMs, on the other hand, may differ pointwise from the Bures metric \cite{Safranek:2017aa}, so the latter may be seen as a ``smoothed'' version of the former.

The goal of this paper is to draw the attention to the role of singular QFIMs, a topic that, with a few significant exceptions~\cite{Safranek:2017aa,Seveso:2019wh}, goes almost unnoticed in quantum metrology. We carry out this task on a case-by-case basis to stress the kind of conundrums that arise in relevant physical situations and give intuition to the possible solutions.

\section{Quantum estimation theory and quantum Fisher information}

Let us start with some basic background~\cite{Helstrom:1976ij}. We consider a classical statistical model~\cite{McCullagh:2002uw} defined by a probability distribution $p(\mathbf{x}| \bmtheta)$, where $\bmtheta  \in \mathbb{R}^{n}$  are the parameters to be estimated and $\mathbf{x} \in \mathbb{R}^{n}$ denotes the outcomes of an experiment performed to deduce the value of  $\bmtheta$. In general, experiments consist of $M$ trials, with outputs typically assumed to be independent and identically distributed. In the following, we take the outcomes to be continuous random variables, although the discrete case can be treated much in the same way.

In order to infer the parameter values $\bmtheta$ we have to build estimators $\widehat{\bmtheta} (\mathbf{x})$; that is, we need a map associating possible sets of measurement outcomes $\mathbf{x}$ to the possible values of the parameters $\bmtheta$. These estimators are said to be unbiased if their mean values coincide with the true values of the unknown parameters (i.e., there is no systematic error in the estimation)~\cite{Kay:1993aa}:
\begin{equation}
\mathbb{E}_{\mathbf{x}} [ \widehat{\bmtheta} ] \equiv \int \D 
\mathbf{x} \, p(\mathbf{x}| \bmtheta) \; \widehat{\bmtheta} (\mathbf{x}) = \bmtheta \, .
\end{equation}
All the estimators we shall consider here are unbiased; they satisfy the famous Cram\'er-Rao bound (CRB)~\cite{Cramer:1946aa,Rao:1945aa}:
\begin{equation}
\label{eq:CRB}
\bm{\Cov}_{\mathbf{x}} ( \widehat{\bmtheta} )  \succcurlyeq  
\frac{1}{M}  \bm{\mathsf{F}}_{\mathbf{x}}^{-1} ( {\bmtheta})  \, ,
\end{equation}
where the inequality $\mathbf{A} \succcurlyeq \mathbf{B}$ means that $\mathbf{A} - \mathbf{B}$ is a positive semidefinite matrix. Here, the covariance matrix  $(i, j = 1, \ldots, n)$
\begin{equation}
[ \Cov_{\mathbf{x}} ( \widehat{\bmtheta} )]_{ij} = \mathbb{E}_{\mathbf{x}} [ ( \widehat{\theta}_{i} - \theta_{i} )  ( \widehat{\theta}_{j} - \theta_{j} ) ]  \, 
\end{equation}
quantifies the sensitivity relative to each parameter, whereas the Fisher information matrix (FIM)~\cite{Fisher:1922wq} $\bm{\mathsf{F}}_{\mathbf{x}} (\bmtheta)$, with elements 
\begin{equation}
[\mathsf{F}_{\mathbf{x}} ( \bmtheta ) ]_{ij}  =   
\int \D \mathbf{x} \;  
p(\mathbf{x} | \bmtheta) \frac{\partial 
\ln p(\mathbf{x}|\bmtheta)}{\partial \theta_{i}} 
\frac{\partial \ln p(\mathbf{x}|\bmtheta)}
{\partial \theta_{j}} \, ,
\label{eq:classical FIM}
\end{equation}
captures the amount of information encoded in the output probabilities.

Note that the CRB requires the FIM to be invertible. When $\bm{\mathsf{F}}_{\mathbf{x}} (\bmtheta)$ is singular, a simple extension of the CRB is~\cite{Stoica:2001aa}
\begin{equation}
\label{eq:modCRB}
\bm{\Cov}_{\mathbf{x}} ( \widehat{\bmtheta} )  \succcurlyeq  
\frac{1}{M}  \bm{\mathsf{F}}_{\mathbf{x}}^{+} ( {\bmtheta})  \, ,
\end{equation}
where now $\bm{\mathsf{F}}_{\mathbf{x}}^{+} ( {\bmtheta})$ denotes the Moore-Penrose pseudoinverse~\cite{Penrose:1955aa,Ben-Israel:1977aa}, although that may constitute an overly optimistic lower bound~\cite{Xavier:2004aa}.  

In the quantum world, we deal with a quantum statistical model, described by a density operator $\varrho$. The parameters $\bmtheta$ are encoded via a quantum channel $\mathcal{E}_{\bmtheta}$, whose action on the state $\varrho$ is the transformation $\mathcal{E}_{\bmtheta} [ \varrho ] = \varrho_{\bmtheta}$~\cite{Chuang:2000aa}. To estimate $\bmtheta$, we perform a measurement that is represented by some positive operator-valued measure (POVM) $\{ \Pi_{\mathbf{x}} \}$~\cite{Holevo:2003fv}. We thus obtain a statistical distribution that, according to Born's rule, is given by $p(\mathbf{x}| \bmtheta) = \Tr ( \varrho_{\bmtheta} \, \Pi_{\mathbf{x}} )$.  The best estimate of $\bmtheta$ can be obtained with the  tools of classical estimation theory, but one has to guarantee the positivity of the quantum state.  Consequently, quantum estimation can be seen as classical estimation, supplemented with the constraints imposed by positivity~\cite{lnp:2004uq}.  

By optimizing over all possible POVMs, that is, over all the possible measurements allowed by quantum mechanics, one obtains the more fundamental quantum Cram\'er-Rao bound (QCRB), which depends exclusively on the quantum statistical model $\varrho_{\bmtheta}$ and can be formulated as
\begin{equation}
\bm{\Cov}_{\varrho} ( \widehat{\bmtheta}) \succcurlyeq  
\frac{1}{M}  \bm{\mathsf{F}}_{\mathbf{x}}^{-1} ( \bmtheta ) 
\succcurlyeq  
\frac{1}{M} 
\bm{\mathsf{Q}}_{\varrho}^{-1} ( \bmtheta ) \, ,
\label{eq:QCRB}
\end{equation}
where now the quantum Fisher information matrix (QFIM) has components~\cite{Petz:2011aa} 
\begin{equation}
[\mathsf{Q}_{\varrho} ( \bmtheta  )]_{ij} =  \tfrac{1}{2} 
\Tr ( \varrho_{\bmtheta} \,  \{ L_{i}, L_{j} \}  ) \,
\end{equation} 
and the subscript $\varrho$ indicates that we are dealing with a quantum model. The operator $L_{i}$ denotes the symmetric logarithmic derivative (SLD) with respect to the parameter $\theta_{i}$ and is implicitly defined by the Lyapunov equation~\cite{Helstrom:1976ij} 
\begin{equation} 
\frac{\partial {\varrho}_{\bmtheta}}{\partial \theta_{i}} = 
\tfrac{1}{2} \{ \varrho_{\bmtheta}, L_{i} \} \, , 
\end{equation}
where $\{ \cdot, \cdot \}$ stands for the anticommutator $\{ A, B \} = A B + B A$. For the particular case of pure states, this reduces to 
\begin{equation}
[ \mathsf{Q}_{\varrho} ( \bmtheta ) ]_{ij} =  
4 \RE \, \langle \partial_{\theta_{i}} \psi_{\bmtheta} | 
\partial_{\theta_{j}} \psi_{\bmtheta} \rangle  + 4 
\langle \partial_{\theta_{i}} \psi_{\bmtheta} | \psi_{\bmtheta} \rangle 
 \langle \partial_{\theta_{j}} \psi_{\bmtheta} | \psi_{\bmtheta} \rangle 
\, .
\end{equation} 

The simultaneous estimation of multiple parameters provides better precision than estimating them individually~\cite{Szczykulska:2016aa}. Unfortunately, attaining the ultimate quantum bounds for such a simultaneous estimation is not always guaranteed~\cite{Yuen:1973aa,Helstrom:1974aa}: the corresponding optimal measurements may not commute, thus making their implementation unrealizable~\cite{Braun:2018aa,Sidhu:2020aa,Albarelli:2020aa,Polino:2020aa,Sidhu:2021tp}. We simply mention that a sufficient condition for the joint estimation is that the operators $L_{i}$ commute.  A weaker condition is provided by the following constraint~\cite{Matsumoto:2002aa}:
\begin{equation}
\label{eq:Matsu}
\Tr ( \varrho_{\bmtheta} [L_{i}, L_{j}]) = 0 \,. 
\end{equation}
For pure states, there exists a necessary and sufficient condition for the saturation of the QCRB: if $\mathsf{Q}_{\varrho} (\bmtheta)$ is invertible,  the QCRB can be saturated if and only if
\begin{equation}
\IM \; \langle \psi_{\bmtheta} |L_{i} L_{j} | \psi_{\bmtheta} \rangle = 0 \, .
\end{equation}

The QFIM is intimately connected with the distinguishability of a probe after undergoing small variations of the parameter~\cite{Wootters:1981aa,Braunstein:1994aa}. The distinguishability between two states, $\varrho_{1}$ and $\varrho_{2}$, can be quantified by the normalized Bures distance~\cite{Bures:1969aa} 
\begin{equation} 
\D_{\mathrm{B}} (\varrho_{1} | \varrho_{2} ) = 
\sqrt{1 - F ( \varrho_{1} | \varrho_{2} )} \, ,
\end{equation} where $F (\varrho_{1} | \varrho_{2} ) = \Tr [ \sqrt{ \sqrt{\varrho_{1}} \, \varrho_{2} \, \sqrt{\varrho_{1}}} ]^{2}$ is the fidelity~\cite{Uhlmann:1976aa}. Then, given two states $\varrho_{\bmtheta}$  and $\varrho_{\bmtheta + \delta \bmtheta}$, obtained by an infinitesimal change in the parameters, one has the Bures metric $\D_{\mathrm{B}}^{2} ( \varrho_{\bmtheta} | \varrho_{\bmtheta + \delta \bmtheta} ) = \sum_{ij} g_{ij} (\bmtheta)  \, \D \theta_{i} \D \theta_{i}$. One can then show that~\cite{Safranek:2017aa} 
\begin{equation} 
\bm{\mathsf{g}} (\bmtheta) = \bm{\mathsf{Q}} (\bmtheta) + 2 \sum_{p(x_{k})=0} 
\bm{\mathsf{H}} (x_{k})  \, ,
\end{equation} 
where $\mathsf{H}_{ij} = \partial^{2} p(\mathbf{x}|\bmtheta)/\partial x_{i} \partial x_{j}$ is the Hessian matrix and must be evaluated over all values $x_{k}$ such that $p(x_{k}|\bmtheta)=0$. Moreover, at the points $x_{k}$,  the QFIM matrix is discontinuous at $\bmtheta$. 

These discontinuities happen  in many instances, which include binomial mixtures, multinomial mixtures, Bayesian networks, neural networks, radial basis functions, hidden Markov models, stochastic context-free grammars, reduced rank regressions, and  Boltzmann machines~\cite{Watanabe:2008ub}. In this context, a quantum statistical model is called \emph{regular} when  it satisfies the following conditions~\cite{Davison:2008vn}: 
\begin{enumerate}[label={C\arabic*.--}]
\item Identifiability. The parametrization $\bmtheta \mapsto \varrho_{\bmtheta}$ is injective. 

\item Fixed rank. The rank of $\varrho_{\bmtheta}$ is independent of $\bmtheta$.
  
\item  Nonsingular metric. The Bures metric is a well-defined function for all $\bmtheta$.
\end{enumerate}
When some of these regularity conditions are not true, the QCRB  does not hold in general.  In a nonidentifiable model, the true value of the parameters is, in general, nonunique~\cite{Stoica:1982vo}. We thus assume that any physical model of interest is identifiable (or can be made so by a suitable reparametrization) and we focus on when the the quantum model does not satisfy C2 and C3.

\section{Tour of singularities in paradigmatic quantum estimation applications}

\subsection{A toy model}

We first consider a simple  model that has been fully discussed in Ref.~\citeonline{Safranek:2017aa}. It is given by the family of states
\begin{equation}
\varrho_{p} = p \ket{0} \bra{0} + (1-p) \ket{1}\bra{1} ,
\end{equation} 
where $\ket{0}$ and $\ket{1}$ form a computational basis. This is a variable-rank statistical model, where the parameter to be estimated $p$ lies within the range $0 \le p \le 1$. It is also a case of single-parameter estimation where the QFIM is simply a $1\times 1$ matrix known as the quantum Fisher information (QFI). For $0 < p < 1$, the QFI reads
\begin{equation}
\mathsf{Q}_{p} = \frac{1}{p (1-p)} \, ,
\end{equation}
which shows a discontinuity in the two limiting values $p = \{0, 1 \}$.   Projecting onto the states $\{ \ket{0} \bra{0}, \ket{1} \bra{1} \}$ determines an optimal measurement and clearly does not depend on the underlying value of the parameter being estimated. For the limiting values $p = \{0, 1 \}$, one can check that a maximum likelihood estimator would give a variance equal to zero, thus violating the QCRB~\cite{Seveso:2019wh} and avoiding the infinite-uncertainty paradox.~\footnote{This contradicts the statement in Ref.~\citeonline{Safranek:2017aa} that ``the quantum Cram\'er-Rao bound holds for the possibly
discontinuous quantum Fisher information matrix''.}

\subsection{Phase estimation}

The paradigmatic example of quantum-enhanced metrology is phase estimation. There, a relative phase that encodes some physically meaningful quantity is imprinted between two arms of an interferometer~\cite{Rafal:2015aa}. Singularities arise also here: some probe states are wholly insensitive to variations in this relative phase. This serves as a launching pad for the many physical scenarios in which probe states cause singular QFIMs by being insensitive to parameters that they purport to sense.

Phase estimation begins with a Hilbert space defining two, often spatial, modes of the electromagnetic field. One mode has an effective path length that is greater than the other, leading to a transformation governed by the unitary operator
\begin{equation}
    {U}_{\mathrm{PE}}(\theta)=\exp\left[\iu \theta ({n}_a - {n}_b )\right],
\end{equation} 
where ${n}_a = {a}^\dagger {a}$ is the number operator for the bosonic mode annihilated by ${a}$ and likewise for the orthogonal mode $b$. Again, this is a single-parameter estimation and the  QFI as a function of $\theta$ for a given probe state ${\varrho}$ is 
\begin{equation}
\mathsf{Q}_{\varrho}(\theta)=4 \, \mathrm{Var}_{\varrho} ({n}_a - {n}_b ) \, .
\end{equation}

It is simple to see that an eigenstate of ${n}_a- {n}_b$ will have vanishing information about the parameter $\theta$; this is equally apparent from such a state being an eigenstate of ${U}_{\mathrm{PE}}$, with only a global phase being imparted on such a state. The defective states for phase estimation are number states $\ket{n_a}\otimes\ket{n_b}$. Physically, this makes perfect sense: each number state has its phase completely undefined in its respective phase space and there are no superpositions present in this separable (i.e., not entangled) state. Here, we witness the confluence of defective states, singular QFIMs, inestimable parameters, and unphysicality of global phases in quantum physics.

At the heart of phase estimation is a relative phase. Is it then surprising that a single-mode state, perhaps with only the vacuum state in the second mode, can have a nonzero QFIM? A canonical coherent state
\begin{equation}
\ket{\alpha} = \eu^{- | \alpha |^{2}/2} \sum_{n=0}^\infty \frac{\alpha^n}{\sqrt{n!}}\ket{n},
\end{equation} 
for example, has QFI 
\begin{equation}
\mathsf{Q}_{\alpha}(\theta)=4 \left|\alpha\right|^2;
\end{equation} 
this is nonzero! This highlights an important characteristic of the QFI, which can sometimes have subtle consequences: it is optimized over all measurement procedures. This means that it implicitly assumes access to unlimited external resources, such as strong phase references against which the relative phases imparted upon the various number states in $\ket{\alpha}$ can be compared. 

The absence of an external reference in phase estimation requires us to treat our probe state ${\varrho}$ as effectively being the phase-averaged state
\begin{equation}
{\varrho} \to \int_0^{2\pi}\frac{d\phi}{2\pi} \eu^{-\iu {N}\phi} \, {\rho} \, 
\eu^{\iu {N}\phi},
\label{eq:phase average no reference}
\end{equation} 
where ${N}= {n}_a+ {n}_b$ is the total photon-number operator. Any single-mode state on its own transforms under this phase averaging into a convex combination of number states, each of which is ineffective for estimating a phase. 
A coherent state, continuing our example, has its QFI transform to
\begin{equation}
\mathsf{Q}_\alpha (\theta)\to \eu^{-|\alpha|^2} \sum_n \frac{|\alpha|^{2n}}{n!}
\mathsf{Q}_{n} (\theta)= 0 \, .
\end{equation}
We thus see another useful implication of singularities in the QFI: they can tell us when additional external resources are required in order to estimate some parameter with a given probe state, such as external phase references.

\subsection{Multiphase estimation}

The potential singularities in phase estimation are equally present in the simultaneous estimation of multiple phases. Multiphase estimation aims to measure $d$ phases relative to some reference phase, or some symmetrized set of $d$ relative phases among $d+1$ arms of an interferometer~\cite{Macchiavello:2003tt,Humphreys:2013wl,Gagatsos:2016vi,Pezze:2017um,Polino:2019aa}. The unitary operator governing this transformation is
\begin{equation}
{U}_{\mathrm{MPE}} (\bmtheta )=\exp \left[\iu 
\sum_{i=1}^d \theta_i ({n}_i - {n}_{0})\right].
\end{equation}

In this case, as for a single phase, separable states of number states $\ket{{n}_0}\otimes\ket{{n}_1}\otimes\cdots\otimes \ket{{n}_d}$ are again eigenstates of the unitary operator and cause the QFIM to completely vanish. This follows from the same physical realization that number states have no well-defined phase and neither do separable combinations of such states. 

The scenario again becomes more intricate when we perform a closer inspection of the role of the reference arm of the interferometer. As in single-phase estimation, the lack of an external phase reference makes estimating all $d+1$ absolute phases impossible.  After undergoing the transformation of Eq.~\eqref{eq:phase average no reference}, now with ${N}=\sum_{i=0}^d {n}_i$, the $d\times d$ QFIM for the $d$ parameters $\bmtheta$ can be computed for any probe state. For example, a probe state comprised of canonical coherent states $\ket{\bm{\upalpha}} \equiv\ket{\alpha_0}\otimes\cdots\otimes \ket{\alpha_d}$ has a QFI matrix with determinant~\cite{Goldberg:2020vw}
\begin{equation}
\det\left[\bm{\mathsf{Q}}_{\bm{\upalpha}}(\bmtheta)\right]=\frac{4^{d}}{\sum_{i=0}^d \left|\alpha_i\right|^2}\prod_{i=0}^d \left|\alpha_i\right|^2.
\end{equation} 
Any input mode having a vacuum state $\alpha_i=0$ causes one fewer phase to be estimable.

In general, when the probe state has the vacuum in the reference mode, the QFIM always becomes singular. The total number of estimable phases always decreases by one for every arm of the interferometer that is in its vacuum state; when $d$ arms are in their vacuum states, we recover the conclusion from single-phase estimation in which nothing can be measured. This is because the operators ${n}_i- {n}_0$ fail to be independent when $\langle {n_0} \rangle =0$ without another reference arm being present, so the total phase $\sum_{i=1}^d\theta_i$ becomes a global phase for each photon-number subspace, thus making one of the parameters $\theta_i$ inaccessible. Singular QFIMs highlight the importance of properly defining and implementing phase references when doing multiphase estimation, which readily generalizes to other estimation scenarios with other essential resources.

\subsection{Rotation measurements}

Another important example of quantum-enhanced metrology is measuring the three parameters of a rotation~\cite{Goldberg:2021uw}. Now, instead of a set of independent phases encoding physically meaningful quantities, the physical parameters are imprinted on a set of interdependent parameters, such as the angle and axis of a rotation. Some of the singularities that arise here are identical in spirit to those arising in phase estimation, while others are only germane to the interdependent nature of the parameters being measured.

We start with the singularities that parallel those of phase estimation. For what probe states is the QFIM always singular? This is easy to answer using an Euler angle parametrization of the rotation:
\begin{equation}
{U}_{\mathrm{rot}} (\Phi,\Theta,\Psi )=\eu^{-\iu\Phi {J}_3} 
\eu^{-\iu\Theta {J}_2}\eu^{-\iu\Psi {J}_3} \, .
\label{eq:Euler angles}
\end{equation}
Here, we have employed angular momentum operators with the usual commutation relations $\left[ {J}_i, {J}_j\right]=\iu\epsilon_{ijk} {J}_k$, $\epsilon_{ijk}$ being the totally antisymmetric unit tensor. Any probe that is an eigenstate of ${J}_3$ will only acquire information about $\Psi$ through its global phase, which is physically inaccessible, so we expect the QFIM to become singular in that case. Then, since Euler angle parametrizations using any set of axes are all relevant, we expect any eigenstate of $\mathbf{n}\cdot\mathbf{J}$ to be insensitive to a rotation parameter and thus have its QFIM be singular. 

This is indeed the case: the QFIM for estimating rotations can always be expressed as~\cite{Goldberg:2021uw}
\begin{equation}
\bm{\mathsf{Q}}_\varrho (\bmtheta ) = 4 \; \mathbf{H}^\top(\bmtheta)  \, 
\bm{\mathsf{C}}_\varrho(\mathbf{J}) \, \mathbf{H}(\bmtheta ) \, ,
\end{equation} 
where $\mathbf{H}(\bmtheta)$ is some matrix describing the choice of parametrization, the superscript $\top$ denotes the transpose, and 
\begin{equation}
\bm{\mathsf{C}}_\varrho (\mathbf{J}) =\frac{1}{2}\int_0^\infty ds \Tr \left( 
[\varrho, \mathbf{J}] \eu^{-\varrho s} [ \mathbf{J}^\top, \varrho ] 
\eu^{-\varrho s} \right)
\end{equation} 
is a state-dependent matrix that for pure states is the symmetrized covariance matrix 
\begin{equation}
[\bm{\mathsf{C}}_\psi (\mathbf{J} ) ]_{ij}= \tfrac{1}{2} 
\langle \{ {J}_i,  {J}_j \} \rangle - \langle {J}_i \rangle \langle {J}_j \rangle .
\end{equation}
To be singular, there must be a basis in which a row and column of $\bm{\mathsf{C}}_\varrho (\mathbf{J})$ vanish. Getting to that basis amounts to redefining the axes of $\mathbf{J}$, so we can inspect the vanishing of the first row and column of $\bm{\mathsf{C}}_\varrho (\mathbf{J})$ without loss of generality. This happens if and only if $[\varrho , {J}_1 ]=0$, which is satisfied by eigenstates of ${J}_1$. The general result is thus that $\bm{\mathsf{C}}_\varrho (\mathbf{J})$ is singular if and only if the probe state is an eigenstate of some angular momentum projection, which in turn ensures that $\bm{\mathsf{Q}}_\varrho (\bmtheta )$ will be singular.

Next, we answer the following question in the affirmative: are there singular QFIMs even for fully functional probe states? This is equivalent to asking whether the matrix $\mathbf{H}( \bmtheta )$ can be singular. Notably, this happens regardless of the probe state, because all of the probe-state dependence comes in the invertibility of $\bm{\mathsf{C}}_\varrho (\mathbf{J})$.

Singular matrices $\mathbf{H}( \bmtheta )$ abound, arising when the coordinate systems defined by the chosen parameterization of $\bmtheta$ are singular. This happens, for example, when the middle Euler angle $\Theta$ vanishes, such that only the sum parameter $\Psi+\Phi$ is estimable while the difference parameter $\Psi-\Phi$ is not. It similarly happens when the angle of rotation vanishes, in which case the axis of rotation cannot be uniquely defined. As we shall see, these singularities are notably alleviated by switching to a new coordinate system in which the particular set of parameters is not longer a degenerate point in the coordinate system.

\subsection{Measurements of loss, absorption, transmission, or reflection}

Loss measurements are at the foundation of spectroscopy and imaging, thus permeating all the branches of science~\cite{Losero:2018tl}. Many optical sensing problems work by determining changes in the intensity of light shone on an object. When that light is in a single mode annihilated by a bosonic operator ${a}$, the total intensity in that mode is given by $\langle {n}_a\rangle$.

After a continuous or discrete evolution, the field can be modelled by the input-output relation 
\begin{equation}
{a}\to \eta \, {a} + \sqrt{1-\eta^2} \, {b} \, ,
\label{eq:input output a loss}
\end{equation} 
where ${b}$ annihilates some bosonic mode orthogonal to ${a}$ (i.e., $[{a},{b}^\dagger ]=0$) that is initially in its vacuum state. This generic measurement problem is now that of estimating the parameter $\eta$. We here encounter a problem that does not involve unitary evolution, so the full machinery of QFI must be used to make progress.

Canonical coherent states are often used in this paradigm. These evolve as
\begin{equation} 
\ket{\alpha} \to \ket{\eta \, \alpha} = \eu^{(1-\eta^2)|\alpha|^2/2} \eta^{{n}_a} \ket{\alpha} \, ,
\end{equation}
furnishing them with the QFI~\cite{Monras:2007wx}
\begin{equation}
\mathsf{Q}_{\alpha} (\eta)=4 \, \mathrm{Var}_{\eta \alpha}( {n}_a/\eta)= 
4 | \alpha |^2 \, .
\label{eq:loss QFI eta coherent}
\end{equation}
As in phase estimation, this trivially becomes singular when $\alpha$ vanishes, because then there is no input field to be absorbed or transmitted. 

Other coordinate systems have more often been used for loss measurements. In Refs.~\citeonline{Monras:2007wx} and \citeonline{Adesso:2009wn}, for example, the loss parameter being considered is defined by $\cos\phi=\eta$, which is intrinsically related to a decay rate in the master equation describing the process. In that case, we find
\begin{equation}
    \mathsf{Q}_{\alpha}(\phi)=4|\alpha|^2\sin^2\phi,
\end{equation} which is singular when $\phi=0$, corresponding to the situation with zero loss. Physically, this means that coherent states cannot be used to estimate a certain loss parameter $\phi$ if the input state is equivalent to the output state.

\subsection{Superresolution of incoherent point sources}

Quite recently, a groundbreaking proposal~\cite{Nair:2016aa, Tsang:2016aa} reexamined what were previously thought to be the fundamental limits to the resolution for any camera. The previous limits assumed the moniker of ``Rayleigh's curse'' because the QFI diverges for estimating the separation between two incoherent sources of light when they get closer to each other than the Rayleigh limit. Remarkably, when considering the problem from the perspective of quantum metrology, the
QFI maintains a fairly constant value for any separation of the
sources, which implies that the Rayleigh criterion is secondary to the
problem at hand and showcases the power of finding true fundamental precision limits.

We consider here the generalization to three incoherent identical sources, with fixed centroid at a known position. This a two-parameter estimation problem with a density matrix given by
\begin{equation}
	\varrho = \frac{1}{3} \left(\ket{\psi_1}\bra{\psi_1}+\ket{\psi_2}\bra{\psi_2}+\ket{\psi_3}\bra{\psi_3}\right) \, .
\end{equation}
where $\ket{\psi_1}=e^{-iP(s+d_1)}\ket{\psi}$, $\ket{\psi_2}=e^{-iP(s+d_2)}\ket{\psi}$, and $\ket{\psi_3}=e^{-iP(s-d_1-d_2)}\ket{\psi}$, with $P$ being the momentum operator, $\ket{\psi}$ the real, symmetric, and smooth point-spread function~\cite{Goodman:2004aa} centered at the origin, $s$ the centroid position, and $d_1$ and $d_2$ the distances from two of the sources to the centroid. We consider the QFIM in the $(d_1,d_2)$-basis. For fixed $d_2$, $\varrho$ is symmetric with respect to $d_1=-d_2/2$, which implies that $\mathsf{Q}_{11}= \mathsf{Q}_{12}=0$ there. These are also points where the density matrix changes its rank, because two of the sources coincide, so the QFIM can develop discontinuities.

The general expression for the QFIM at all points for which the density matrix does not change its rank can be calculated using the method proposed in Refs.~\citeonline{Safranek:2018aa} and \citeonline{Fiderer:2021ue}. We obtain 
\begin{eqnarray}
 \mathsf{Q}_{11}(d_1,d_2) & = & \frac{2}{3}\left[4\alpha+\frac{\beta^2_{21}(4-\gamma_{12}^2)+\beta_{23}^2(4-\gamma_{23}^2)+2\beta_{21}\beta_{23}(\gamma_{12}\gamma_{23}+2\gamma_{13})}{\gamma_{12}^2+\gamma_{13}^2+\gamma_{23}^2+\gamma_{12}\gamma_{13}\gamma_{23}-4}\right] , \nonumber
 \\
 \mathsf{Q}_{22}(d_1,d_2) & = &\mathsf{Q}_{11}(d_2,d_1) , 
 \label{eq:QFIM superresolution}
 \\
 \mathsf{Q}_{12}(d_1,d_2) & = & \frac{2}{3} \left\{ 2\alpha -\frac{\beta_{21}^2(\gamma_{12}^2-4)+\beta_{13}\beta_{23}(2\gamma_{12}+\gamma_{13}\gamma_{23})}
 	{\gamma_{12}^2+\gamma_{13}^2+\gamma_{23}^2+\gamma_{12}\gamma_{13}\gamma_{23}-4} \right. \nonumber \\
 & & \qquad \qquad \left. \frac{+\beta_{21} \left[\beta_{13}(2\gamma_{23}+\gamma_{12}\gamma_{13})-\beta_{23}(2\gamma_{13}+\gamma_{12}\gamma_{23})\right]}
	{\gamma_{12}^2+\gamma_{13}^2+\gamma_{23}^2+\gamma_{12}\gamma_{13}\gamma_{23}-4} \right\} , \nonumber
\end{eqnarray}
with $\alpha=\bra{\psi}P^2\ket{\psi}$, $\gamma_{ij}=\Braket{\psi_i|\psi_j}$ and $\iu \beta_{ij}=\bra{\psi_i}P\ket{\psi_j}$. The above expressions for  $\mathsf{Q}_{11}$ and  $\mathsf{Q}_{12}$ do not tend to zero when we approach the line $d_1=-d_2/2$, so they are only valid in the regions where $d_1\neq-d_2/2$ and we conclude that the QFIM must be discontinuous. This is in contrast to the Bures metric, which can be continuous when the QFIM is not \cite{Safranek:2017aa} and is equal to the expressions in Eq. \eqref{eq:QFIM superresolution} even when $d_1=-d_2/2$, so the QFIM differs from the Bures metric along the line $d_1=-d_2/2$.

Again due to the symmetry of $\varrho$ with respect to $d_1=-d_2/2$ for fixed $d_2$, $\mathsf{F}_{11}$ and $\mathsf{F}_{12}$ will vanish when $d_1=-d_2/2$. In contrast to $\mathsf{Q}_{11}$ and $\mathsf{Q}_{12}$, which avoid going to zero  because of their discontinuous behaviour, we expect the classical FIM calculated using Eq. \eqref{eq:classical FIM} to tend to zero in the region $d_1\approx -d_2/2$. To make the FIM equal to the QFIM we must produce a discontinuity for all $d_1=-d_2/2$. Is it possible to find such a classical FIM that exactly equals the QFIM, including the discontinuous behaviour of the latter?

The way to force a discontinuity in $\mathsf{F}_{11}$ and $\mathsf{F}_{12}$, so that they can be compatible with $\mathsf{Q}_{11}$ and $\mathsf{Q}_{12}$,  is by making a measurement (e.g., projecting onto the PSF derivatives~\cite{Rehacek:2017aa,Rehacek:2018vq}) that produces probabilities that go to zero, yielding a discontinuous classical FIM.  However, in contrast to the two-source case, where a discontinuity at one point is required, in the three-source case a discontinuity for a continuum of points is required. Trying a strategy similar to the one in Ref.~\citeonline{Rehacek:2018vq} fails: it is possible to produce a discontinuity at the origin ($d_1=0,d_2=0$) but not for the continuum of points $d_1=-d_2/2$. It looks like there does not exist any direct POVM that will achieve the QCRB for this problem. However, because the PSF and its SLDs can all be taken to be real, the condition of Eq. \eqref{eq:Matsu} is met. This may imply that the saturability of the multiparameter QCRB is even more nuanced when the density matrix has parameter-dependent rank.
Note, regardless, that a substantial improvement is achieved by measuring a POVM made up of the PSF and its derivatives over direct detection, even if the quantum limit cannot be reached, as the ultimate limit achievable is always within a factor of order unity from the QCRB~\cite{Carollo:2019va}.

Summing up, superresolution seems to be related to the discontinuous behaviour of the QFIM, which avoids Rayleigh's curse by producing discontinuities. 

\section{Strategies for taming singularities}

Refreshingly, singular QFIMs are often tameable. Different strategies are superior in different physical scenarios, so we continue exploring them through the relevant examples discussed above.

\subsection{Changing measurement protocols}

There are many situations in which the FIM is singular. A prominent example is that of Rayleigh's curse, in which the FIM for estimating the separation between two incoherent sources of light by measuring the intensity distribution in the image plane is quadratic in this separation, making the FIM singular in the limit that the separation approaches zero. Importantly, a measurement protocol that also incorporates the phase information of the light is not singular there. This was a strong example that demonstrated the importance of designing appropriate measurement protocols to shatter previous limitations~\cite{Paur:2016aa,Yang:2016aa,Tham:2016aa}.

Changing measurement protocols will never rectify singular QFIMs. The QFIM is optimized over all measurement protocols for a given probe state, meaning that it is insensitive to changes in said protocols. We will hence turn our focus to situations in which the QFIM itself is singular, for which strategies other than improving the measurement protocol are necessary to proceed.

\subsection{Changing parametrization}

Many singular QFIMs can be rectified by changing the relationships between the parameters being measured. In proceeding, we stress that a change in parametrization differs from a change in basis, as the latter has no effect on the invertibility of the QFIM.

We first consider the case of rotations. We saw above that rotation measurements can be singular when $\mathbf{H}(\bmtheta )$ is singular. For the Euler angles defined by Eq.~\eqref{eq:Euler angles}, we find
\begin{equation}
\mathbf{H} (\Phi,\Theta,\Psi) = 
\begin{pmatrix}
0&0&1\\
-\sin\Phi&\cos\Phi&0\\
\sin\Theta\cos\Phi& \sin\Theta\sin\Phi &\cos\Theta
\end{pmatrix} \, ,
\qquad \det \mathbf{H}(\Phi,\Theta,\Psi) = - \sin\Theta \, ;
\end{equation}
as alluded to, one cannot estimate the difference $\Phi-\Psi$ when $\Theta\to 0$. Now, if we switch coordinate systems, such as defining a different set of Euler angles
\begin{equation}
{U}_{\mathrm{rot}}(\alpha,\beta,\gamma)=\eu^{-\iu \alpha {J}_1} \eu^{-\iu \beta\op{J}_2}\eu^{-\iu \gamma\op{J}_3},
\end{equation} we instead find
\begin{equation}
\mathbf{H}(\alpha,\beta,\gamma) = 
\begin{pmatrix}
1&0&0\\
0&\cos\alpha&\sin\alpha\\
\sin\beta&-\sin\alpha\cos\beta&\cos\alpha\sin\beta
\end{pmatrix} \, , \qquad 
\det \mathbf{H}(\alpha,\beta,\gamma)=\cos\beta \, .
\end{equation} 
This new coordinate system indeed has a singularity, when the middle Euler angle is a quarter revolution, because that serves to interchange two of the coordinate axes. Notably, these singularities are different from the singularity of the $(\Phi,\Theta,\Psi)$-parametrization: the former is invertible for the case of zero rotation and the latter is singular! This trend continues to hold for other parametrizations of rotations, such as using the angle and axis of rotation, which demonstrates how singular QFIMs can signify undefined coordinate systems and be rectified by switching to a coordinate system that is nondegenerate there. The method of changing parametrizations is mathematical, not physical, so can be used in a variety of physical settings without significant effort.

For loss measurement, we saw above that it can have singular QFIMs for particular 
parameters values $\phi=0$ when using a coherent-state probe. Since this is a singularity that disappears for other values of $\phi$, can it be rectified by a coordinate change? Absolutely: using the original $\eta$ parameter yielded the nonsingular QFI of Eq. \eqref{eq:loss QFI eta coherent}: $\mathsf{Q}_\alpha(\eta)=4|\alpha|^2$. 

In this new parametrization, all of the parameter dependence has vanished from the QFI. By simply viewing the problem as estimating $\eta$ from Eq. \eqref{eq:input output a loss}, we have completely alleviated the singularity in the QFI. This shows how a judicious choice of parameters is always useful and, moreover, that an unfortunate choice of parameters can be easily rectified. 


The optimal states for absorption measurements are Fock states $\ket{n}$, which have definite photon number~\cite{Adesso:2009wn}. These transform as
\begin{equation}
\ket{n}\bra{n} \to \sum_{k=0}^n p_k \ket{k}\bra{k}, \qquad p_k=\binom{n}{k}\cos^{2k} \phi \sin^{2n-2k}\phi,
\end{equation} 
which are the first mixed states we have encountered for computing the QFI:
\begin{equation}
\mathsf{Q}_n (\phi)=\sum_{k=0}^n \frac{1}{p_k}\left(\frac{\partial p_k}{\partial \phi}\right)^2 = 4n \qquad\Rightarrow\qquad\mathsf{Q}_n(\eta)=\frac{4n}{1-\eta^2}\, .
\end{equation}
For a fixed intensity, these outperform canonical coherent states by a factor of $\sin^2\phi$, which significantly helps in the extreme of zero loss. By switching the probe state used for sensing, the singularities in the QFI are avoided here, a topic to which we will return.

These examples lead to another interpretation of the parametrization of the QFIM. The QFIM tells us about curvature in the space of probability distributions induced by a given quantum probe state, with different parametrizations scaling the coordinates in different ways. Because different states have parameter-independent QFIMs for different parametrizations, we conclude that some parametrizations make the metric flat for particular probe states. This means that there are parametrizations whose effect on the curvature of the probability distributions exactly cancels the curvature induced by particular probe states.

\subsection{Changing weights: Weight by metric}

There is another way to vanquish coordinate singularities without resorting to a change parametrization: one can choose a weight matrix that best changes the QCRB to a scalar inequality. Inspecting Eq. \eqref{eq:QCRB}, we introduce a matrix $\mathbf{W}\succ 0$ to find a scalar bound for the weighted mean-square error
\begin{equation}
    \mathrm{wMSE}( \widehat{\bmtheta}) \equiv \Tr [\mathbf{W} \bm{\Cov}_{\varrho} ( \widehat{\bmtheta}) ] \geq \frac{1}{M} \Tr [\mathbf{W} \bm{\mathsf{Q}}_{\varrho}^{-1} ( \bmtheta ) ]\equiv C_S(\bmtheta) \, .
\end{equation} 
The quantity $\mathrm{wMSE}( \widehat{\bmtheta})$ is called the weighted mean square error of the estimator. This scalar inequality is essential for providing a single figure of merit to optimize in multiparameter estimation scenarios.

How should the weight matrix $\mathbf{W}$ be chosen? When the different parameters depend on each other, it is not at all obvious how to assign relative weights to each parameter. This problem was solved for a specific case in Ref.~\citeonline{Rafal:2020aa} and in general in Ref.~\citeonline{Goldberg:2021vj}: choose $\mathbf{W}$ to be the Cartan metric tensor for the geometry associated with the evolution~\cite{Helgason:1978vb}. The Cartan metric defines an intrinsic set of coordinates that exists regardless of the initial parametrization. Given this choice, all of the parameter dependence in the QCRB exactly cancels out and we are left with a scalar QCRB that encapsulates the essential characteristics of a probe state stripped of parametrization-choice problems.

A straightforward example is again the case of rotation measurements. We saw above that the QFIM is singular in various Euler-angle parametrizations for various choices of parameters. Fortunately, these singularities are \textit{exactly} canceled by the properly-chosen weight matrix, regardless of the parametrization. If we choose the weight matrix to correspond to the metric tensor, it cancels the singularity $\det \mathbf{H}=-\sin\Theta$ by having $\det\mathbf{W}=\csc^2\Theta$ in the $(\Phi,\Theta,\Psi)$-coordinates; simultaneously, it cancels the singularity $\det\mathbf{H}=\cos\beta$ by having $\det\mathbf{W}=\sec^2\beta$ in the $(\alpha,\beta,\gamma)$-coordinates. These properties and more follow because $\mathbf{H}^\top \mathbf{H}$ is parametrization-independent and proportional to the metric tensor. It is then apparent that choosing 
\begin{equation}
    \mathbf{W}=\mathbf{H}^{\top} \mathbf{H}
\end{equation} 
yields the scalar bound for rotation estimation
\begin{equation}
    \mathrm{wMSE}(\widehat{\bmtheta})\geq \frac{1}{4M} \Tr [ \bm{\mathsf{C}}_\varrho^{-1}(\mathbf{J}) ] \, .
    \label{eq:scalar rotation bound}
\end{equation} 
Since $\bm{\mathsf{C}}_\varrho(\mathbf{J} )$ depends only on the physical problem at hand and the probe state being used, this intrinsic choice of weight matrix yields a QCRB that captures the usefulness of a probe state for a given measurement scenario.

The example for rotations holds true for other physical scenarios, most prominent among which are all measurement paradigms where the probe state gains information about the parameters to be measured through a unitary evolution. These allow us to essentially find a QFIM similar to $\bm{\mathsf{C}}_\varrho(\mathbf{J} )$ that depends only on the probe state and the geometry of the evolution without depending on fickle choices of parametrization.

We can use this protocol even when the QFIM is already a scalar, as in single-parameter estimation. In many scenarios, such as phase estimation, the QCRB is already parameter independent in the most common parametrization. This is because many single-parameter estimation protocols involve estimating the phase $\theta$ from some generic unitary, such as  ${U}(\theta)=\exp (\iu \theta {G} )$, 
for which the QFI is $\theta$-independent:
\begin{equation}
\mathsf{Q}_\varrho(\theta) = 4\, \mathrm{Var}_\varrho( {G}) \, ,
\qquad 
C_S(\theta)   = \frac{1}{4M} \frac{1}{\mathrm{Var}_\varrho({G})} \, .
\end{equation} 
One could equally well consider some slightly more involved unitary operator ${U}(\theta)= \exp [\iu f(\theta) \, {G} ]$,  where now we may discover singularities in the QFI
\begin{equation}
\mathsf{Q}_\varrho \left[f(\theta)\right] = 4 \, \left[\frac{\partial f(\theta)}{\partial \theta}\right]^{-2} \mathrm{Var}_\varrho( {G})
\end{equation}
depending on the function $f(\theta)$. The Cartan metric tensor will exactly cancel the Jacobian transforming from the parameter $\theta$ to $f(\theta)$ in the QFI, so  $\mathrm{wMSE} [\widehat{f(\theta)} ]= \mathrm{Var}_\varrho (\theta)$
and the intrinsic bound exactly matches the $\theta$-bound 
\begin{equation}
\mathrm{wMSE}\left[\widehat{f(\theta)}\right]\geq \frac{1}{4M}  \frac{1}{\mathrm{Var}_\varrho( {G}) } \, .
\end{equation}

Perplexingly, there is no weight matrix that makes the QFI for loss estimation to depend only on the probe state and not the underlying parameters. This is most easily seen because different parametrizations yield a  parameter-independent QCRB for different probe states. Why is this the case? The answer can be gleaned from Eq. \eqref{eq:input output a loss}: while this evolution equation looks unitary, the QFI involves inferring the loss parameter from only one part of the evolved global state. The corresponding metric, on the other hand, deals with the entire unitary evolution, about which the measured mode has only partial information. Since the vacuum mode is not measured, singularities arise when all of the parameter information is contained in that vacuum mode instead of in the measured mode. Access to the auxiliary modes involved in the problem would allow one to obtain a parameter-independent QCRB that holds regardless of probe state; fundamentally, this would also allow one to achieve Heisenberg-scaling precision in estimating the loss parameter \cite{Monras:2007wx}. We thus see how singular QFIMs can inform us about our ignorance to certain information arising from the nature of our probe states, which is an important situation that cannot be rectified even with an ideal choice of weight matrix.

\subsection{Changing probe states}

Different parametrizations and weight matrices are useful to avoid singularities but, sometimes, even they are insufficient. For example, the parametrization-independent scalar QCRB for rotation measurements found in Eq. \eqref{eq:scalar rotation bound} can have singular $\bm{\mathsf{C}}_\varrho(\mathbf{J} )$ for particular probe states $\varrho$. We saw above that parametrization-independent rotation-measurement singularities arise if and only if the probe state is an eigenstate of an angular momentum projection operator. This directly implies that, for a given physical problem, certain probe states must be avoided if one wishes to gain full access to all of the underlying parameters.

The trick is to discover why certain probe states are insensitive to the underlying parameters being measured. In the case of rotation measurements, the states garnering singular QFIMs are those that are unchanged after undergoing rotations about a particular axis. In the case of multiphase estimation, the states garnering singular QFIMs are those that are unchanged after the application of a particular phase. These examples may seem to depend on all of the parameters being applied, but their dependence is through a global phase, which is fundamentally immeasurable. 

There are other physical scenarios in which a probe state does not depend on the parameter in question. This is most easily seen using absurd examples: in most experiments, the states being prepared will not have any of their properties depend on, say, the colour of a planet on the far side of the universe. We can readily rule out using probe states that obviously do not have any dependence on a parameter in question. 

Why is quantum metrology any different? Why can we easily rule out absurd probe states in some scenarios and not others? The answer tends to be that quantum mechanics can be more subtle, so that one can accidentally imprint a parameter as a global phase. It is the evolution $\exp(-\iu \Psi {J}_3)\ket{\psi}=\exp(-\iu\Psi j_3)\ket{\psi}$ that explains why a thought-to-be sensitive state is truly insensitive to changes in some parameter $\Psi$. In the Euler-angle parametrization of Eq.~\eqref{eq:Euler angles}, for example, it is apparent that an eigenstate of ${J}_3$ will only depend on $\Psi$ through an immeasurable global phase, but it is far from obvious that an eigenstate of ${J}_1$ will be insensitive to changes in some function of $\Phi$, $\Theta$, and $\Psi$. One must always ensure that their probe states are independently sensitive to changes in all of the underlying parameters for the former to be globally useful; otherwise, singular QFIMs abound. The hallmark of needing a change in probe state is that the QFIM remains singular for all changes in parametrization. Then, if the QFIM is always singular even after changing probe states, one must conclude that none of their states actually depend on the parameters in question.

\section{Concluding remarks}

We have presented  a number of physical case studies that underline the role of singular and discontinuous QFIMs. Singularities appear when the probe states, coordinate systems, or parameters being estimated are defective and discontinuities appear when varying estimation parameters changes the rank of the density matrix describing a quantum state.  In some instances, these problems  can be a source of great confusion, as it it is not usual to see singular or discontinuous functions in physics. The QCRB breaks down in these pathological cases, sometimes in both the classical and quantum case. As a consequence, special care has to be paid in these situations.

\section*{Acknowledgments}
JLR, ASS, and LLSS acknowledge financial support from the European Union’s Horizon 2020 research and innovation program (Projects ApresSF and Stormytune). LLSS acknowledges funding from the Spanish Ministerio de Ciencia e Innovación (Grant No. PGC2018- 099183-B-I00). 


\begin{thebibliography}{10}

\bibitem{Helstrom:1976ij}
C.~W. Helstrom, {\em Quantum {D}etection and {E}stimation {T}heory} (Academic,
  New York, 1976).

\bibitem{Tsang:2020ve}
M.~Tsang, F.~Albarelli and A.~Datta, {\em Phys. Rev. X} {\bf 10} (2020)  031023.

\bibitem{Rafal:2020aa}
R.~Demkowicz-Dobrza{\'n}ski, W.~G{\'o}recki and M.~Gu{\c t}{\u a}, {\em J.
  Phys. A: Math. Theor.} {\bf 53}  (2020)  363001.

\bibitem{Goldberg:2021uw}
A.~Z. Goldberg, A.~B. Klimov, G.~Leuchs and L.~L. S{\'{a}}nchez-Soto, {\em J.
  Phys: Photonics} {\bf 3}  (2021) 022008.

\bibitem{Monras:2007wx}
A.~Monras and M.~G.~A. Paris, {\em Phys. Rev. Lett.} {\bf 98}  (2007) 
  160401.

\bibitem{Tsang:2019ww}
M.~Tsang, {\em Contemp. Phys.} {\bf 60} (2019) 279.

\bibitem{Goldberg:2020vw}
A.~Z. Goldberg, I.~Gianani, M.~Barbieri, F.~Sciarrino, A.~M. Steinberg and
  N.~Spagnolo, {\em Phys. Rev. A} {\bf 102} (2020) 022230.

\bibitem{Bures:1969aa}
D.~Bures, {\em Trans. Am. Math. Soc.} {\bf 135}  (1969) 199.

\bibitem{Safranek:2017aa}
D.~\v{S}afr\'anek, {\em Phys. Rev. A} {\bf 95}  (2017) 052320.

\bibitem{Seveso:2019wh}
L.~Seveso, F.~Albarelli, M.~G. Genoni and M.~G.~A. Paris, {\em J. Phys. A:
  Math. Theor.} {\bf 53}  (2019) 02LT01.

\bibitem{McCullagh:2002uw}
P.~McCullagh, {\em Ann. Stat.} {\bf 30} (2002) 1225.

\bibitem{Kay:1993aa}
S.~M. Kay, {\em Fundamentals of Statistical Signal Processing} (Prentice Hall,
  Upper Saddle River, 1993).

\bibitem{Cramer:1946aa}
H.~Cram{\'e}r, {\em Mathematical Methods of Statistics} (Princeton University
  Press, Princeton, NJ, 1946).

\bibitem{Rao:1945aa}
R.~C. Rao, {\em Bull. Calcutta Math. Soc.} {\bf 37} (1945) 81.

\bibitem{Fisher:1922wq}
R.~A. Fisher and E.~J. Russell, {\em Philos. T. Roy. Soc. A} {\bf 222}  (1922)
  309.

\bibitem{Stoica:2001aa}
P.~{Stoica} and T.~L. {Marzetta}, {\em IEEE Trans. Signal Process.} {\bf 49}
  (2001) 87.

\bibitem{Penrose:1955aa}
R.~Penrose, {\em Proc. Cambridge Phil. Soc.} {\bf 51}  (1955) 406.

\bibitem{Ben-Israel:1977aa}
A.~Ben-Israel and T.~N.~E. Greville, {\em Generalized Inverses: Theory and
  Applications} (Wiley, New York, 1977).

\bibitem{Xavier:2004aa}
J.~{Xavier} and V.~{Barroso}, The {R}iemannian geometry of certain parameter
  estimation problems with singular {F}isher information matrices, in {\em Int.
  Conf. Acoust. SPEE\/},  (Montreal, 2004).

\bibitem{Chuang:2000aa}
I.~Chuang and M.~Nielsen, {\em Quantum {C}omputation and {Q}uantum
  {I}nformation} (Cambridge University Press, Cambridge, 2000).

\bibitem{Holevo:2003fv}
A.~S. Holevo, {\em Probabilistic and {S}tatistical {A}spects of {Q}uantum
  {T}heory}, 2nd edn. (North Holland, Amsterdam, 2003).

\bibitem{lnp:2004uq}
M.~G.~A. Paris and J.~\v{R}eh\'a\v{c}ek (eds.), {\em Quantum State Estimation},
  Lect. Not. Phys., Vol.~649 (Springer, Berlin, 2004).

\bibitem{Petz:2011aa}
D.~Petz and C.~Ghinea, {\em Introduction to {Q}uantum {F}isher {I}nformation},
  in {\em Quantum Probability and Related Topics\/},  (World Scientific, 2011),
  pp. 261--281.

\bibitem{Szczykulska:2016aa}
M.~Szczykulska, T.~Baumgratz and A.~Datta, {\em Adv. Phys. X} {\bf 1} (2016)
  621.

\bibitem{Yuen:1973aa}
H.~Yuen and M.~Lax, {\em IEEE Trans. Inf. Theory} {\bf 19}  (1973) 740.

\bibitem{Helstrom:1974aa}
C.~Helstrom and R.~Kennedy, {\em IEEE Trans. Inf. Theory} {\bf 20}  (1974) 16.

\bibitem{Braun:2018aa}
D.~Braun, G.~Adesso, F.~Benatti, R.~Floreanini, U.~Marzolino, M.~W. Mitchell
  and S.~Pirandola, {\em Rev. Mod. Phys.} {\bf 90}  (2018) 035006.

\bibitem{Sidhu:2020aa}
J.~S. Sidhu and P.~Kok, {\em AVS Quantum Sci.} {\bf 2}  (2020) 014701.

\bibitem{Albarelli:2020aa}
F.~Albarelli, M.~Barbieri, M.~G. Genoni and I.~Gianani, {\em Phys. Lett. A}
  {\bf 384}  (2020) 126311.

\bibitem{Polino:2020aa}
E.~Polino, M.~Valeri, N.~Spagnolo and F.~Sciarrino, {\em AVS Quantum Sci.} {\bf
  2}  (2020) 024703.

\bibitem{Sidhu:2021tp}
J.~S. Sidhu, Y.~Ouyang, E.~T. Campbell and P.~Kok, {\em Phys. Rev. X} {\bf
  11} (2021) 011028.

\bibitem{Matsumoto:2002aa}
K.~Matsumoto, {\em J. Phys. A: Math. Gen.} {\bf 35}  (2002) 3111.

\bibitem{Wootters:1981aa}
W.~K. Wootters, {\em Phys. Rev. D} {\bf 23}  (1981) 357.

\bibitem{Braunstein:1994aa}
S.~L. Braunstein and C.~M. Caves, {\em Phys. Rev. Lett.} {\bf 72}  (1994) 3439.

\bibitem{Uhlmann:1976aa}
A.~Uhlmann, {\em Rep. Math. Phys.} {\bf 9}  (1976) 273.

\bibitem{Watanabe:2008ub}
S.~Watanabe, {\em Algebraic geometrical method in singular statistical
  estimation}, in {\em Quantum {Bio-Informatics}\/},  (World Scientific, Singapore, 2008),
  pp. 325--336.

\bibitem{Davison:2008vn}
A.~C. Davison, {\em Statistical Models} (Cambridge University Press, Cambridge,
  2008).

\bibitem{Stoica:1982vo}
P.~Stoica and T.~S{\"o}derstr{\"o}m, {\em Int. J. Contr.} {\bf 36} (1982)
  323.

\bibitem{Rafal:2015aa}
R.~Demkowicz-Dobrza{\'n}ski, M.~Jarzyna and J.~Ko{\l}ody{\'n}ski, {\em Quantum
  Limits in Optical Interferometry}, in {\em Progress in Optics\/},  ed.
  E.~Wolf (Elsevier, 2015), pp. 345--435.

\bibitem{Macchiavello:2003tt}
C.~Macchiavello, {\em Phys. Rev. A} {\bf 67} (2003) 062302.

\bibitem{Humphreys:2013wl}
P.~C. Humphreys, M.~Barbieri, A.~Datta and I.~A. Walmsley, {\em Phys. Rev.
  Lett.} {\bf 111} (2013) 070403.

\bibitem{Gagatsos:2016vi}
C.~N. Gagatsos, D.~Branford and A.~Datta, {\em Phys. Rev. A} {\bf 94} (2016)
  042342.

\bibitem{Pezze:2017um}
L.~Pezz{\`e}, M.~A. Ciampini, N.~Spagnolo, P.~C. Humphreys, A.~Datta, I.~A.
  Walmsley, M.~Barbieri, F.~Sciarrino and A.~Smerzi, {\em Phys. Rev. Lett.}
  {\bf 119} (2017) 130504.

\bibitem{Polino:2019aa}
E.~Polino, M.~Riva, M.~Valeri, R.~Silvestri, G.~Corrielli, A.~Crespi,
  N.~Spagnolo, R.~Osellame and F.~Sciarrino, {\em Optica} {\bf 6}  (2019) 288.

\bibitem{Losero:2018tl}
E.~Losero, I.~Ruo-Berchera, A.~Meda, A.~Avella and M.~Genovese, {\em Sci. Rep.}
  {\bf 8}  (2018) 7431.

\bibitem{Adesso:2009wn}
G.~Adesso, F.~Dell'Anno, S.~De~Siena, F.~Illuminati and L.~A.~M. Souza, {\em
  Phys. Rev. A} {\bf 79}  (2009) 040305.

\bibitem{Nair:2016aa}
R.~Nair and M.~Tsang, {\em Phys. Rev. Lett.} {\bf 117} (2016) 190801.

\bibitem{Tsang:2016aa}
M.~Tsang, R.~Nair and X.-M. Lu, {\em Phys. Rev. X} {\bf 6} (2016) 031033.

\bibitem{Goodman:2004aa}
J.~W. Goodman, {\em Introduction to {F}ourier {O}ptics} (Roberts and Company,
  Englewood, 2004).

\bibitem{Safranek:2018aa}
D.~\v{S}afr\'anek, {\em Phys. Rev. A} {\bf 97}  (2018) 042322.

\bibitem{Fiderer:2021ue}
L.~J. Fiderer, T.~Tufarelli, S.~Piano and G.~Adesso, {\em PRX Quantum} {\bf 2}
  (2021) 020308.

\bibitem{Rehacek:2017aa}
J.~Rehacek, M.~Pa{\'u}r, B.~Stoklasa, Z.~Hradil and L.~L. S{\'a}nchez-Soto,
  {\em Opt. Lett.} {\bf 42}  (2017) 231.

\bibitem{Rehacek:2018vq}
J.~{\v R}eh{\'a}{\v c}ek, Z.~Hradil, D.~Koutn{\'y}, J.~Grover, A.~Krzic and
  L.~L. S{\'a}nchez-Soto, {\em Phys. Rev. A} {\bf 98} (2018) 012103.

\bibitem{Carollo:2019va}
A.~Carollo, B.~Spagnolo, A.~Dubkov and D.~Valenti, {\em J. Stat. Mech. Theory
  Expt.} {\bf 2019}  (2019) 094010.

\bibitem{Paur:2016aa}
M.~Paur, B.~Stoklasa, Z.~Hradil, L.~L. Sanchez-Soto and J.~Rehacek, {\em
  Optica} {\bf 3}  (2016) 1144.

\bibitem{Yang:2016aa}
F.~Yang, A.~Taschilina, E.~S. Moiseev, C.~Simon and A.~I. Lvovsky, {\em Optica}
  {\bf 3}  (2016) 1148.

\bibitem{Tham:2016aa}
W.~K. Tham, H.~Ferretti and A.~M. Steinberg, {\em Phys. Rev. Lett.} {\bf 118}
  (2016) 070801.

\bibitem{Goldberg:2021vj}
A.~Z. Goldberg, L.~L. S\'anchez-Soto and H.~Ferretti, {\em Phys. Rev. Lett.}
  {\bf 127}  (2021)

\bibitem{Helgason:1978vb}
S.~Helgason, {\em Differential geometry, {L}ie groups, and symmetric spaces}
  (Academic, New York, 1978).

\end{thebibliography}

\end{document}